# ChartDesign: Towards LLM Designer of Data Visualization

**Mohammed Afaan Ansari**
University of Maryland, College Park
College Park, MD, USA
mansari1@umd.edu

**Aniruddh Bansal**
University of Maryland, College Park
College Park, MD, USA
ani01@umd.edu

**Tianyi Zhou**
University of Maryland, College Park
College Park, MD, USA
tianyi.david.zhou@gmail.com

## Abstract

Charts are the dominant medium for visualizing data, discovering patterns and trends, and communicating data-driven insights, yet designing them still requires expensive human effort and expertise, such as selecting appropriate chart types, axis orientations, font sizes, and layouts. Most automatic visualization systems rely on handcrafted heuristics or simple rule matching and therefore struggle to generalize across domains. This work explores the potential of large language models (LLMs) as *chart designers*. We propose ChartDesign, which post-trains LLMs to imitate human experts and generate chart design attributes given tabular data. To this end, we curate a diverse training corpus of data–design pairs from charts in public surveys (PewResearch) and academic repositories (CharXiV). Vision–language models are used to extract data and design attributes from these charts, including chart type, sub-type, alignment, titles, axis labels, and bar spacing, formatted as JSON. We then fine-tune LoRA adapters on Phi-3, Qwen-3, and InternVL2.5 to learn a mapping from data to design specifications. ChartDesign significantly improves chart design performance over strong baselines, achieving up to 84% accuracy on a held-out test set (vs. 53% for the best baseline) and generalizing to unseen domains. We further show that charts rendered from ChartDesign-generated specifications are visually appealing and human-preferred, narrowing the human–AI gap in data visualization.

## 1 Introduction

Data visualizations are indispensable tools for analysts, scientists, and end-users. A well-designed chart not only summarizes quantitative trends, reveals insightful patterns, but also communicates subtle relationships that might be missed by textual descriptions alone. Creating such charts, however, remains an expensive manual process. For example, the analyst must choose an appropriate chart type (bar, line, pie, etc.), decide whether bars should be grouped or stacked, align axes and legends, tune grid lines and spacing, and compose concise labels. While libraries such as `matplotlib` and `ggplot2` expose rich design controls, most automatic plotting tools rely on generic defaults, forcing novices into suboptimal designs and experts into repetitive manual tuning.

Can we train an LLM to act as a *chart designer*? Recent work has demonstrated that VLMs are adept at recognizing chart images and extracting the underlying data Kim et al. (2023); Liu et al. (2023), yet generating a useful chart specification from scratch remains under-explored. Straightforward prompting of off-the-shelf LLMs yields generic JSON templates or chart types misaligned with human preference because these models lack sufficient understanding of visualization principles. At the same time, classical design-recommendation engines such as Data2Vis Dibia & Demiralp (2019) and AdaVis Ma et al. (2021) rely on handcrafted grammar rules and thus fail to generalize beyond a handful of chart styles.

In this paper, we post-train LLMs to **learn to design data visualizations** by imitating human-authored charts. Our post-trained model, `ChartDesign`, takes a data table as input and predicts a structured `JSON` description of the chart design. The key insight is to train LLMs to produce human-preferred design choices; for example, a line chart is more appropriate for time-series data than a pie chart, along with fine-grained layout parameters such as axis titles, bar spacing, and grid lines, etc. To this end, we curate a novel dataset of chart designs from two complementary sources: (i) *Pew Research*Kantharaj et al. (2022) surveys, which provide polished infographics spanning public opinion, economics and culture, and (ii) *CharXiV*Wang et al. (2024), an archive of academic plots extracted from arXiv papers in computer science, physics and statistics. We pair each image with its underlying data (extracted via a VLM) and manually annotate the chart's design attributes using a comprehensive JSON schema. The resulting corpus comprises 2118 charts covering bar, line, area, scatter and pie plots as well as grouped and stacked variants.

We then finetune lightweight LLMs to map the raw data table to this JSON representation. Parameter-efficient LoRA adapters enable us to train models as small as 4 billion parameters (Phi-3-mini-4k-instruct) and 8 billion parameters (Qwen-3) on modest hardware. During inference the same model can produce a bespoke chart specification for a new data table without requiring the original image. Figures 2, 3, 4 provide an overview of the dataset construction, training, and evaluation pipeline.

This paper makes the following contributions:

- We introduce **`ChartDesign`**, a framework for post-training LLMs to act as data-visualization designers. Unlike prior work focused on data extraction or library-specific code generation Dibia & Demiralp (2019); Ma et al. (2021); Tian et al. (2024), our model outputs a *renderer-agnostic* design specification (a structured JSON) capturing chart type, orientation, labels and spacing, which can be rendered across Matplotlib, Vega-Lite, Altair and ggplot2 without modification.
- We curate and release a diverse dataset of 2,118 human-authored charts from PewResearch and CharXiV, each paired with extracted CSV data and a rich design JSON. The held-out test set (100 Pew + 100 CharXiV) is *fully human-verified*, providing a reliable benchmark for chart design evaluation.
- We demonstrate that parameter-efficient LoRA fine-tuning substantially improves design prediction, achieving 20–30 percentage-point gains over strong zero-shot and few-shot baselines, with further gains from full fine-tuning. An LLM judge validated against human raters (92% agreement on 120 held-out pairs) confirms the reliability of our automatic evaluation.
- We present expanded human evaluation (10 annotators, 40 chart pairs) showing that 65% of annotators prefer `ChartDesign`-generated designs, alongside error analysis and robustness experiments that characterise the remaining failure modes.

## 2 Related Work

**Chart understanding and extraction.** Recent years have witnessed significant progress in automatic chart understanding, encompassing chart image classification, data extraction Wang et al. (2020) and structure parsing Chen et al. (2022). Vision-language models such as LLaVA Liu et al. (2023) and ChartOCR Chen et al. (2022) can recognize axes and legends and even output the underlying data values. These methods are complementary to our work: we assume the data table is available as input and instead focus on predicting high-level design choices.

**Visualization design recommendation.** Visualization recommendation systems fall into two broad categories. Early rule-based approaches, such as APT Mackinlay (1986a), SAGE Roth et al. (1994), ShowMe Mackinlay et al. (2007) and Voyager Wongsuphasawat et al. (2016), encode design heuristics derived from perceptual studies and rank possible visual encodings for a data table. While these systems provide interpretable recommendations, they require extensive hand-crafted rules and do not scale well to the combinatorial design space. Learning-based methods instead infer design patterns directly from data. Examples include Data2Vis Dibia & Demiralp (2019), VisML Watanabe et al. (2022), AdaVis Ma et al. (2021) and subsequent extensions that translate tables into Vega-Lite or similar specifications.

Table 1: Comparison with representative prior systems along three key dimensions.

| System | Output | Training | Evaluation |
|---|---|---|---|
| Data2Vis Dibia & Demiralp (2019) | Vega-Lite spec | Seq2Seq | Spec correctness |
| AdaVis Ma et al. (2021) | Vega-Lite spec | Rule + ML | Rec. accuracy |
| ChartGPT Tian et al. (2024) | Plot code | Prompting | User preference |
| VisText Tang et al. (2023) | Caption | Seq2Seq | NLG metrics |
| **`ChartDesign`** | **Design JSON** | **LoRA / Full FT** | **Attr. accuracy + human** |

**LLMs for chart generation.** Several recent systems leverage large language models to produce chart code from natural-language descriptions. Approaches such as ChartGPT Tian et al. (2024), NL4DV-LLM Narechania et al. (2020) and Chat2Vis Maddigan & Susnjak (2023) decompose a user query into prompts and then instruct an LLM to emit Vega-Lite or Python plotting code. These agentic pipelines are tied to a specific rendering engine, often require iterative feedback, and aim to map natural-language intent to executable code. Because they focus on code generation, they do not explicitly supervise perceptual design attributes. In contrast, our method trains LLMs to output a language-independent design specification: a structured JSON describing chart type, orientation, labels and spacing. This separation of design from rendering allows the same specification to be implemented in multiple plotting libraries and enables the model to internalise general design principles.

**Datasets for chart understanding and design.** Several datasets pair chart images with structured annotations, but differ substantially in task framing from our work. VisText Tang et al. (2023) provides scene-graph representations and source tables for chart captioning; its scene graphs overlap with our design attributes, but the task is natural-language generation rather than design prediction. ChartQA Masry et al. (2022) targets question answering over charts and does not provide renderer-agnostic design JSONs. FigureQA Kahou et al. (2018) and DVQA Kafle et al. (2018) are synthetically generated and lack the stylistic diversity of human-authored charts. Closest to our setting is the CharXiV Wang et al. (2024) benchmark for chart understanding, which we repurpose as a source of academic design examples. Unlike all of the above, `ChartDesign` provides paired (CSV, design JSON) training data and a human-verified evaluation set specifically targeting the *design prediction* problem, i.e. inferring the visual specification that a human expert would choose for a given data table.

Table 1 summarises how `ChartDesign` differs from the most closely related systems along three key dimensions.

## 3 ChartDesign

`ChartDesign` is a training recipe that equips language models with a notion of visualization design. In this section we describe the curation of our dataset, our training pipeline using parameter efficient LoRA adapters, and the key observations from our ablation studies. We emphasise the major contributions of our work: (i) the creation of a diverse dataset of data-design pairs, (ii) a loss computation and sampling strategy tailored to imbalanced design attributes, and (iii) an LLM-based evaluation framework for fine-grained assessment.

### 3.1 Training data curation

**Source corpus.** Our dataset combines 2,118 charts drawn from two complementary repositories. We sample 1,101 polished infographics from PewResearch Pew Research Center (2015); Masry et al. (2022) and 1,017 scientific plots from CharXiV Wang et al. (2024). Together they span bar, line, area, scatter and pie charts, include grouped and stacked variants, and cover both categorical and temporal domains. The mixture of presentation-oriented and academic styles exposes the model to diverse

| Statistic | Pew | CharXiV |
|---|---|---|
| Total charts | 1,101 | 1,017 |
| Bar (primary) | 600 | 200 |
| Line (primary) | 400 | 420 |
| Pie (primary) | 50 | 20 |
| Area (primary) | 30 | 110 |
| Scatter (primary) | 0 | 180 |
| Box (primary) | 0 | 70 |
| Histogram (primary) | 0 | 17 |
| Grouped bar | 50 | 35 |
| Stacked bar | 20 | 15 |
| Alignment (values) | 3 | 3 |
| H. grid lines | 0–4 | 0–6 |

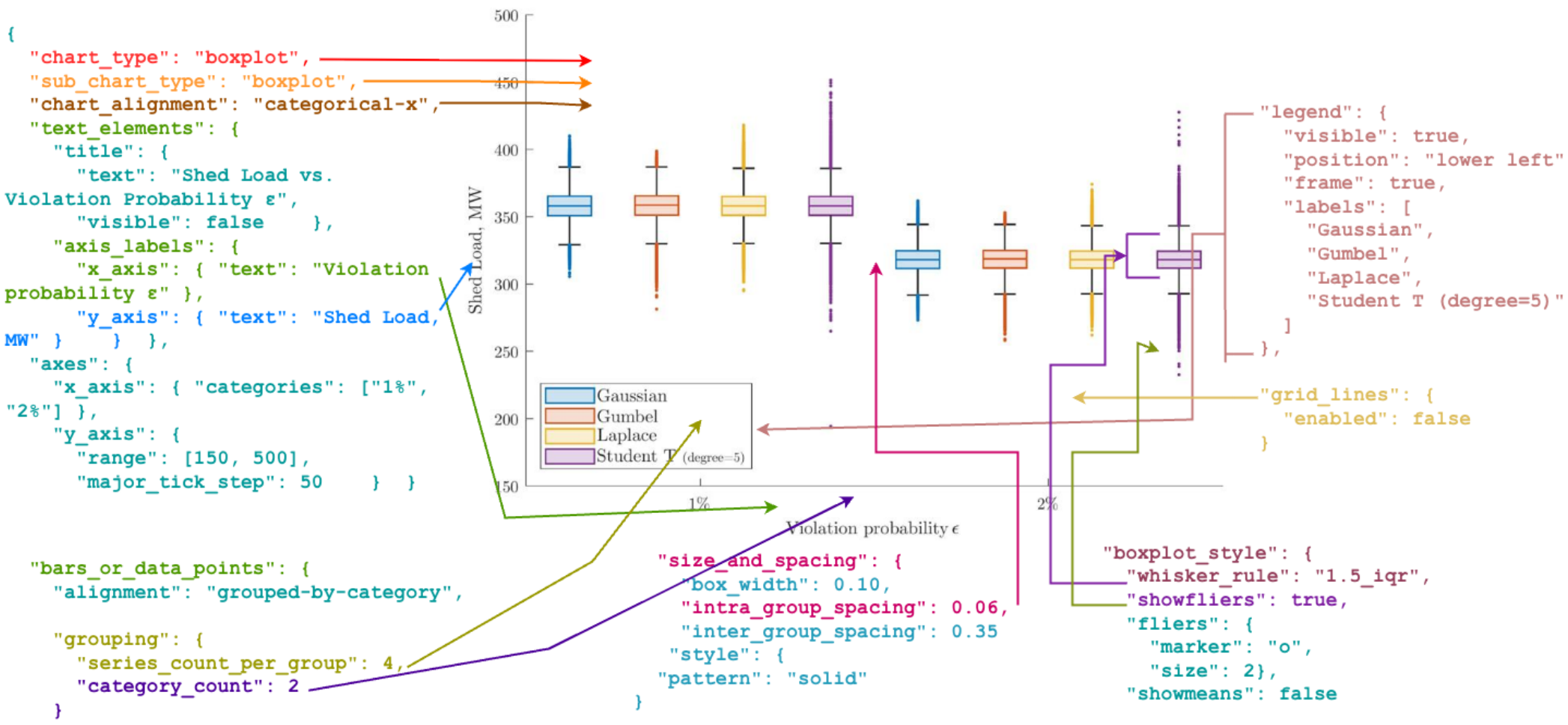

Figure 1: **Design schema overview.** A sample box plot (right) annotated with selected fields from our design JSON (left). Arrows indicate how visual components map to schema attributes such as `chart_type`, `text_elements`, `axes`, `legend`, `bars_or_data_points` and `boxplot_style`.

design decisions. Additional statistics are provided in Appendix A.4.

**Data extraction.** We extract the underlying numerical data for each chart by prompting an off-the-shelf vision-language model (`Phi-3.5-Vision Instruct`) to output one or more CSV tables. The chat prompt instructs the model to return plain CSV with column headers and to label multiple tables as “Chart 1:”, “Chart 2:”, etc. This simple protocol lets us handle multi-chart images without introducing unstructured explanations. We manually verify and correct the extracted CSVs; if a parse is erroneous we re-generate or discard it to ensure clean training data. An overview of the extraction pipeline is provided in Figure 2.

**Design annotation.** Each chart is paired with a hierarchical JSON capturing its salient design choices. Inspired by perceptual effectiveness guidelines Cleveland & McGill (1985); Mackinlay (1986b), the schema includes fields for the chart type and sub-type, the overall orientation, free-form text elements, counts of horizontal and vertical grid lines, the alignment of graphical marks, their width and spacing, and a categorical style pattern (solid, striped or dotted). Fields that do not apply are omitted so that the schema generalises across chart families. These attributes correspond to parameters exposed by common plotting libraries and reflect factors known to affect readability and interpretability. For example, a simple line chart would have `chart_type line`, a horizontal `chart_alignment`, and `text_elements` storing its title and axis labels; a bar chart specification would additionally include bar width, spacing and pattern. Figure 1 illustrates how a sample box plot is annotated with selected fields from our schema. A complete specification of the schema and annotation guidelines is provided in Appendix A.5.

**Human verification and test sets.** As the dataset is generated via LLM annotation, we conduct manual verification on a subset of the examples to ensure label fidelity. Specifically, we sample 100 PewResearch charts and 100 CharXiV charts as a held-out test set and manually validate both the extracted CSVs and the corresponding design JSONs. This human-annotated subset serves as a high-quality benchmark for evaluating our models and is kept separate from the training data.

### 3.2 Training pipeline

**Model architecture.** We use the chat-oriented Phi-3-mini-4k-instruct (4.2B parameters) and Qwen-3 (8B) models, both of which support long context windows suitable for encoding CSV tables.

**Parameter-efficient finetuning.** To adapt the base models with manageable memory requirements, we employ LoRA adapters: we freeze the original weights, insert rank-16 projections in attention and feed-forward layers, and train only these adapters. Additional finetuning details, such as GPU configuration and batch sizing, are provided in Appendix A.2.

**Input representation.** Each training example consists of a user instruction and assistant response. The instruction includes a high-level task description ("Generate a JSON for chart attributes") followed by the CSV data. The assistant response is the annotated JSON. We wrap these messages in the chat template of the respective model. As some tables are long, we set the maximum length to eight thousand tokens for Qwen and fourteen thousand for Phi-3.

**Loss and sampling.** The models are trained to minimise the auto-regressive cross-entropy loss between the predicted tokens and the ground-truth JSON. A naïve uniform sampling of training examples would overfit to common chart types (e.g., vertical bar charts) and underperform on minority classes (e.g., scatter plots or striped patterns). To mitigate this imbalance we compute per-attribute inverse frequency weights and sample training batches accordingly. Let $n_k$ denote the number of training samples containing attribute value $k$ and $N$ be the total number of samples. We define an importance weight for $k$ as $w_k = \log\big(1 + N/n_k\big)$. The sampling weight for a training example is then the product of the weights of all attribute values present in its JSON (with a small penalty for missing attributes), normalised across the dataset. In practice we draw batches of size four using a *WeightedRandomSampler* (PyTorch) with replacement. This ensures that underrepresented classes such as “scatter” or “striped” are seen more frequently during training. During training we monitor the sampled distribution and log class frequencies for each batch to verify balanced coverage. We experimented with directly weighting the token-level loss but found that modifying the sampling distribution yields more stable optimisation.

**Optimisation and hyperparameters.** Models are fine-tuned for three epochs on 90% of the dataset and validated on the remaining 10%, using AdamW with learning rate $10^{-4}$ and mixed-precision training. Hyper-parameter settings and ablation results are deferred to Appendix A.2.

### 3.3 Evaluation

**Metrics.** We evaluate models using *per-attribute accuracy* on a human-verified test set and report an *overall accuracy* computed as the macro-average across attributes Following Sharma et al. (2023). Because design attributes vary in form (categorical labels, free-text strings, or numeric values), exact string matching is insufficient. We therefore use an LLM-based judge to determine semantic equivalence between predicted and ground-truth attribute values (e.g., “bar” vs. “bar chart”).To mitigate potential bias introduced by LLM-based evaluation, we complement this automatic assessment with a human evaluation of rendered charts (§4.3), which directly measures perceptual fidelity and user preference. Full evaluation details are provided in Appendix A.3.

**Baselines and ablations.** To contextualise our results we compare against off-the-shelf vision-language models (LLaVA, Gemma, etc.) prompted with the same instruction but without finetuning. These models struggle to infer correct chart designs, achieving only 24-55 % average accuracy on our test sets. We further ablate the effect of the weighted sampler by training with uniform sampling; performance drops by 5-10 percentage points on rare attributes, confirming the importance of balancing the training distribution. Additional ablations examine the impact of LoRA rank and training data size; details are reported in Appendix A.6.

**LLM-judge validation.** To assess the reliability of our automatic evaluation, we hand-labelled 120 attribute pairs drawn from the test set and compared the LLM judge’s MATCH/NO-MATCH decisions against three human-expert raters. The LLM judge agreed with the majority human label in 92% of cases (8% disagreement, typically due to numeric-tolerance ambiguities or synonym variation). This agreement confirms that the judge is a reliable proxy for human scoring. The held-out test sets (100 Pew + 100 CharXiV) used for all quantitative results are *fully human-verified*, so label noise in the automatically annotated training data does not affect evaluation integrity.

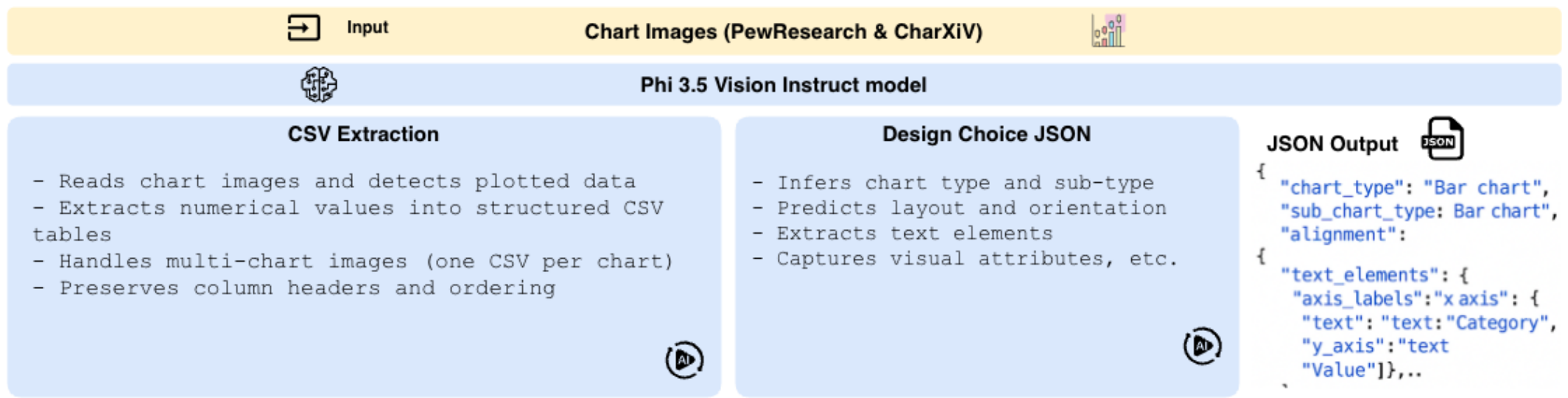

Figure 2: **Dataset construction and annotation pipeline.** Chart images are processed by a vision-language model to extract the underlying data as CSV tables and infer design attributes as a structured JSON. Prompt templates are provided in Appendix A.1.

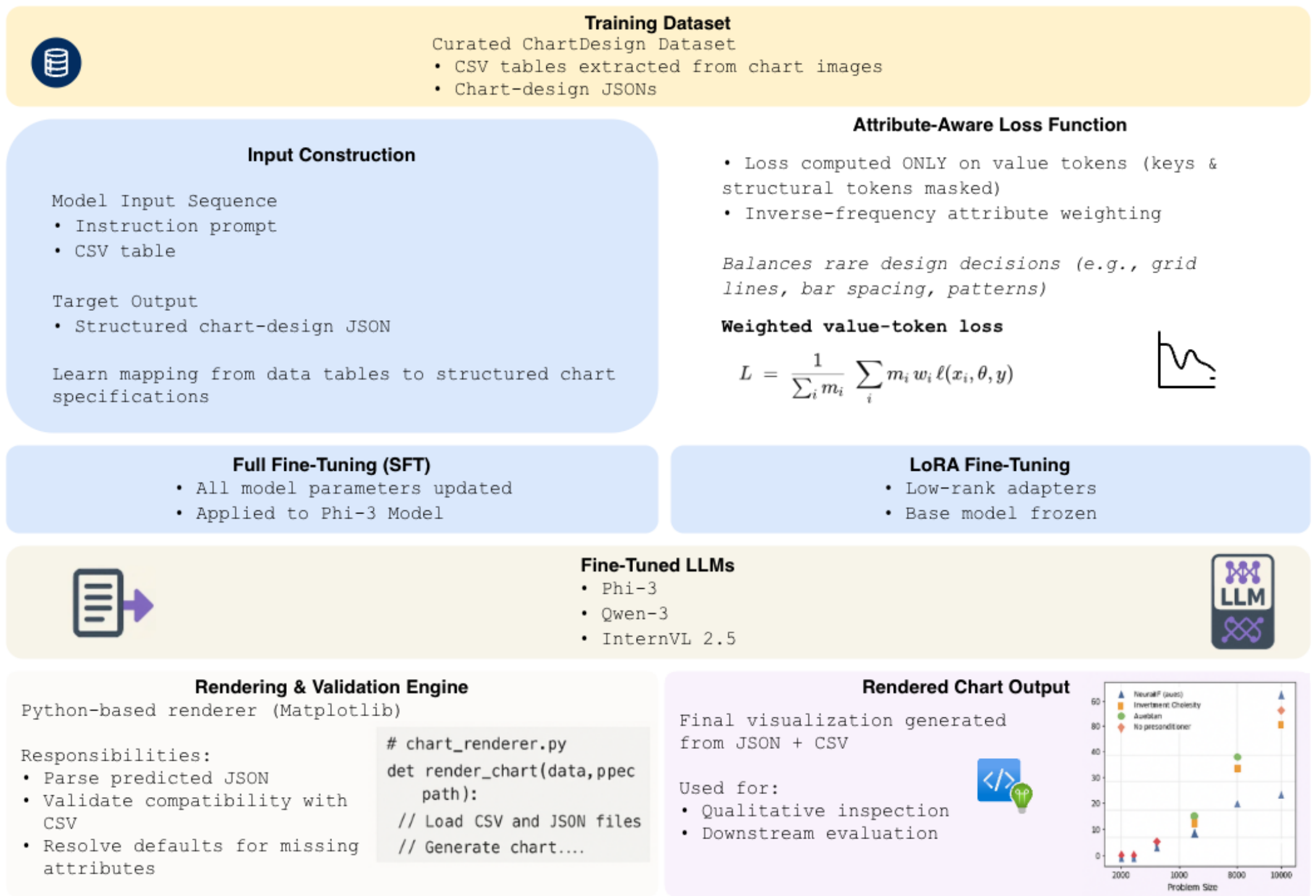

Figure 3: **Model fine-tuning pipeline.** Starting from a curated dataset of CSV tables and design JSONs, a chart LLM is trained to map tabular data and instructions to structured design specifications using an attribute-aware loss with inverse-frequency weighting. The base model is adapted via supervised fine-tuning or LoRA adapters to obtain specialised chart LLMs.

## 4 Experiments

We evaluate `ChartDesign` on two held-out test sets and compare against a range of baselines. Our evaluation focuses on two complementary aspects: (i) the ability of the models to recover the correct design attributes from a data table, measured at the attribute level and in aggregate, and (ii) the perceptual quality of the charts rendered from the predicted specifications.

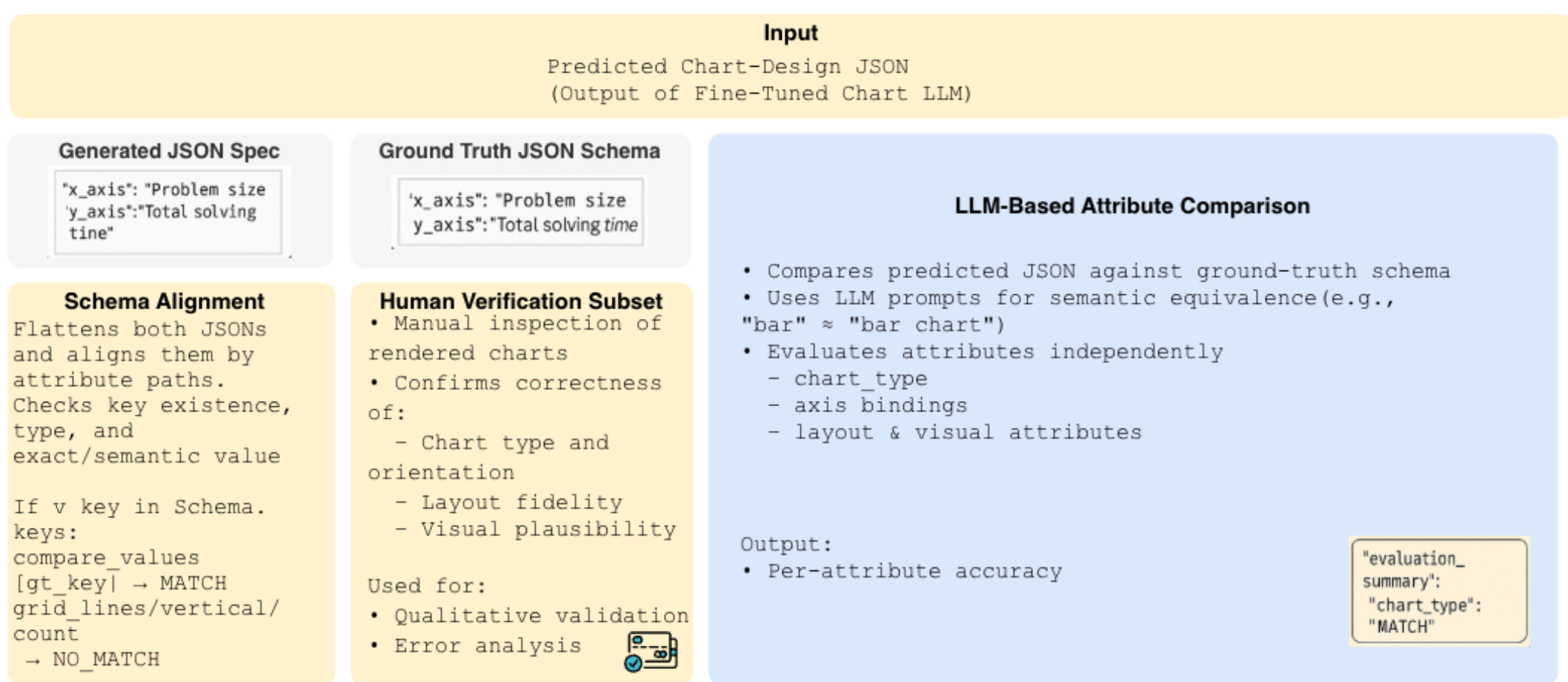

Figure 4: **LLM-based evaluation pipeline.** Predictions are compared to ground truth by flattening the JSONs and using an LLM judge to determine semantic equivalence for each attribute. Matches are aggregated into accuracy metrics.

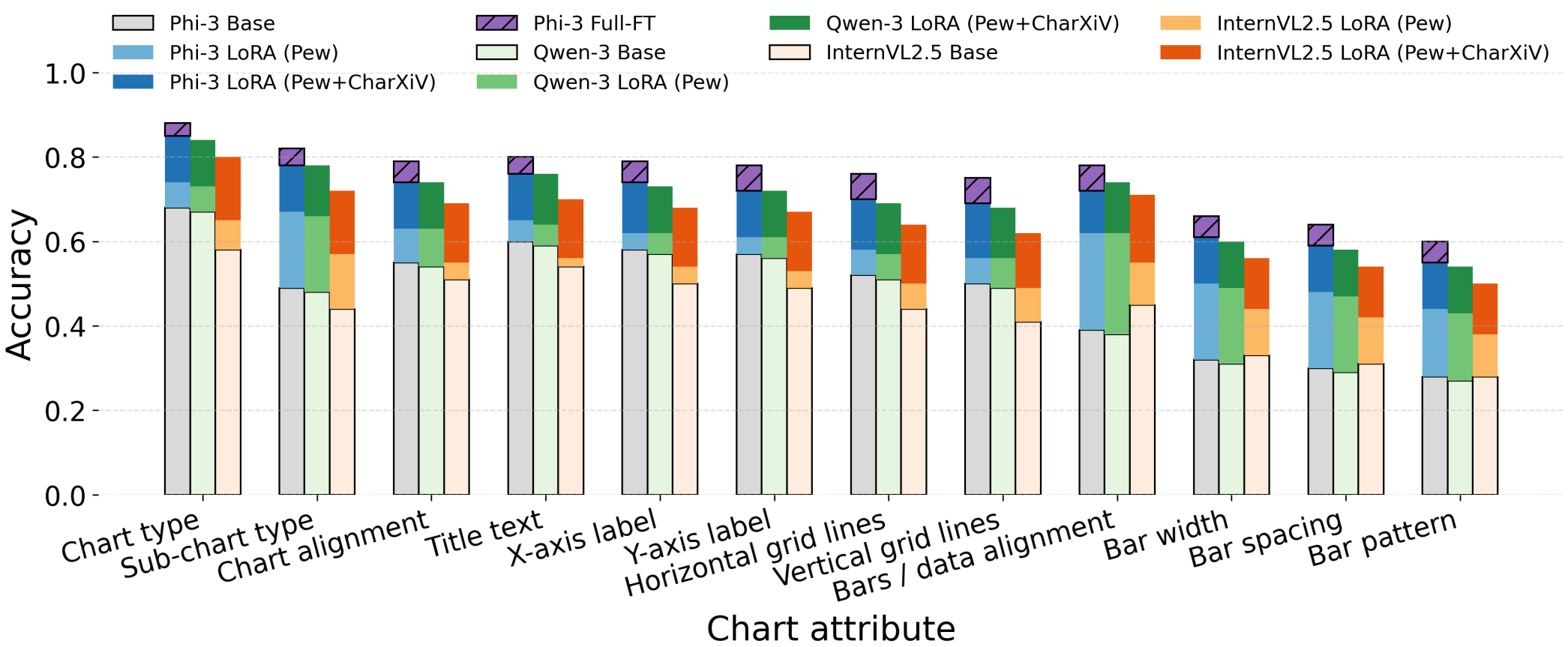

Figure 5: **Attribute-wise accuracy on the PewResearch + CharXiV test set.** Comparison of base models, LoRA fine-tuning, and full fine-tuning across different chart attributes on the mixed-domain evaluation set.

### 4.1 Experimental setup

We evaluate three training variants - **Base** (no fine-tuning), **Finetuned (Pew)** (1,101 PewResearch charts), and **Finetuned (Pew+CharXiV)** (2,118 charts) - applied to Phi-3 (4B), Qwen-3 (8B), and InternVL 2.5 MPO 8B. Test sets are 100 PewResearch charts (in-domain) and 100 Pew + 100 CharXiV charts (cross-domain), both human-verified; we report macro-averaged per-attribute accuracy. Baselines include InternVL 2.5 MPO 8B, Gemma 3 27B, DeepSeek Janus Pro 7B, Phi-3.5 Vision 4.2B, LLaVA 1.6 (Mistral/Vicuna), LLaVA 1.5 7B, Llama 3.2 Vision 11B, and PixTral 12B, all prompted without fine-tuning; off-the-shelf models perform poorly (Table 3), motivating task-specific supervision.

### 4.2 Main results

**Quantitative results.** Table 3 summarises results. Without fine-tuning, Phi-3/Qwen-3 achieve 52-53% on PewResearch and 31-32% on the mixed domain. LoRA fine-tuning yields 72-75% (Pew) and 60-62% (mixed); combined-corpus LoRA reaches up to 84% (Qwen-3, Pew). Full fine-tuning of Phi-3 achieves the highest accuracy (88% Pew, 82% mixed). Predicted JSONs render faithfully across Matplotlib, Vega-Lite, Altair, and ggplot2 (Figure 9).

Table 3: Accuracy (%) on the PewResearch (Pew) and Pew+CharXiV test sets. *Zero-shot*: no fine-tuning. *Few-shot (3-shot)*: three in-context examples. `ChartDesign`: same backbone after LoRA or Full FT under different adaptation regimes.

| **Model** | **Pew ↑** | **Pew+CharXiV ↑** |
|---|---|---|
| *Zero-shot base models* | | |
| InternLM-2.5 (8B) | 32.0 | 24.0 |
| InternLM-3 (8B) | 35.0 | 28.0 |
| Gemma (7B) | 40.0 | 28.0 |
| DeepSeek-LLM (7B) | 35.0 | 25.0 |
| Phi-3 (4B) | 38.0 | 26.0 |
| Ministral-3 (8B) | 42.0 | 30.0 |
| Llama 3 (8B) | 43.0 | 30.0 |
| Llama 3.2 (11B) | 44.0 | 32.0 |
| Mistral (7B) | 41.0 | 29.0 |
| *Zero-shot backbone (pre-fine-tuning baseline)* | | |
| Phi-3 (4B) | 53.0 | 32.0 |
| Qwen-3 (8B) | 52.0 | 31.0 |
| InternLM-2.5 (8B) | 32.0 | 24.0 |
| *Few-shot prompting (3-shot in-context learning)* | | |
| Qwen-3 (8B) | ≈51.0 | ≈46.0 |
| `ChartDesign` *(LoRA on Pew)* | | |
| Phi-3 | 74.0 | 60.0 |
| Qwen-3 | 72.0 | 61.0 |
| InternVL-2.5 | 75.0 | 62.0 |
| `ChartDesign` *(LoRA on Pew+CharXiV)* | | |
| Phi-3 | 78.0 | 72.0 |
| Qwen-3 | 84.0 | 80.0 |
| InternVL-2.5 | 80.0 | 70.0 |
| `ChartDesign` *(Full FT on Pew+CharXiV)* | | |
| Phi-3 | **88.0** | **82.0** |

**Attribute-wise analysis.** Figure 7 plots the accuracy for each design attribute on the PewResearch test set. The bar charts contrast the LoRA-tuned Phi-3 model with its fully fine-tuned counterpart. While both models improve substantially over the untuned backbone, full fine-tuning yields a further boost of roughly ten percentage points across most attributes, especially for rare decisions such as bar width, spacing and pattern. Figure 5 shows the corresponding results on the mixed test set; here full fine-tuning again improves the Phi-3 model without degrading its performance on PewResearch charts.

**Qualitative comparisons.** To assess perceptual quality, we render the predicted JSONs and compare them against the original charts and the base models (Phi-3). As illustrated in Figure 6, the base model often misidentifies the chart type or misaligns axes, leading to misleading visuals. The fine-tuned model reliably reproduces the layout and label structure of the original design. In particular, our model places titles and axis labels correctly, selects appropriate marker shapes and bar orientations, and matches grid line counts.

**Ablation study.** We ablate sampling strategy and training data composition; results are summarized in Appendix A.6.

### 4.3 Human evaluation

Ten annotators each evaluated 40 chart pairs across bar, line, scatter, area, box, and pie charts from both domains. Predicted charts achieve 94% chart-type correctness, 90% orientation correctness, layout fidelity 4.4/5, and visual plausibility 4.6/5. Notably, 65% of annotators preferred `ChartDesign`-generated charts over the original (25%), with 10% expressing no preference-indicating human-competitive quality. Full protocol, results, and figures are in Appendix A.7 (Table 5, Figure 8).

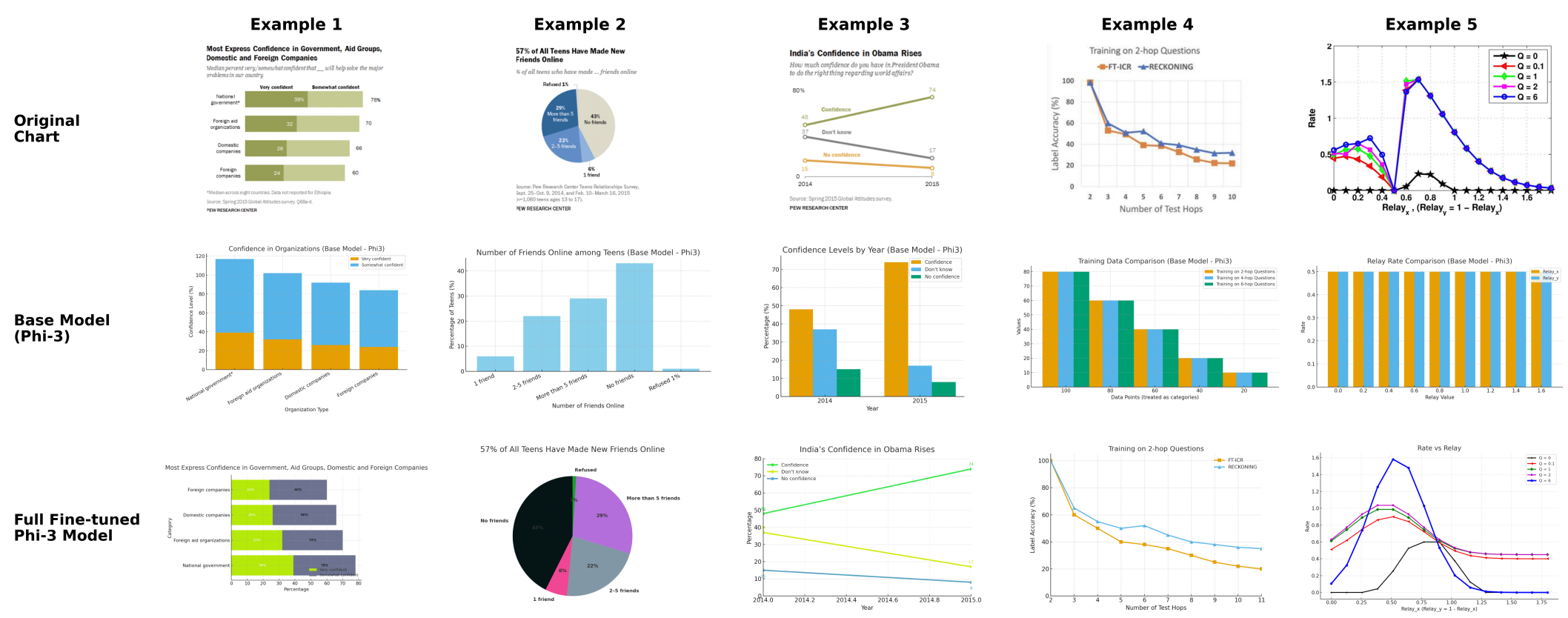

Figure 6: **Qualitative comparison across 5 charts.** Each example shows the original PewResearch chart, the output from a base Phi-3 model, and our finetuned model (Qwen-3). We observe that the base model often misidentifies chart types and alignment, while the finetuned model closely matches the target layout and encoding.

### 4.4 Findings and discussion

The quantitative results above reveal several insights into chart design modelling.

**Finding 1**

Post-training on human authored charts with JSON-formatted design choices significantly improves LLMs' capabilities as chart designers.

Fine-tuning yields 25–35 point gains; sub-chart type improves from 49% to 75% and bar alignment from 39% to 70%. Domain transfer is strong: Pew-only training generalises to CharXiV, with a further 7–10 point gain from mixed training. Detailed attribute-wise analysis is in Appendix A.5.

**Finding 2**

The difficulty of learning to make design choices for different visualization attributes varies.

**Attribute-wise discussion.** Detailed analysis is provided in Appendix A.5. These trends are consistent with perceptual studies on graphical encoding effectiveness Cleveland & McGill (1985); Mackinlay (1986b).

**Finding 3**

Limitations of supervised finetuning (imitation) and future directions.

Extended analyses covering rare-type generalisation, error taxonomy, robustness to noisy inputs, and a data-vs-metadata masking ablation are provided in Appendix A.11-A.14.

## 5 Limitations

Limitations of scope, data quality, evaluation scale, and model coverage are discussed in Appendix A.15.

## 6 Conclusions

We introduced `ChartDesign`, a framework for training large language models to play the role of data-visualization designers. By curating a diverse corpus of human-authored charts, annotating them with a rich hierarchy of design attributes and adapting compact LLMs via parameter-efficient LoRA adapters, we enabled the models to map data tables directly to visualization specifications. Our models significantly outperform pretrained baselines and produce human-preferred charts. Future work includes extending the design schema to additional chart families and exploring interactive refinement through natural language feedback.

## References

Yun Chen, Zhen Mei, and Zhihao Guo. Chartocr: End-to-end chart data extraction. *IEEE Transactions on Pattern Analysis and Machine Intelligence*, 2022. 2

William S. Cleveland and Robert McGill. Graphical perception and graphical methods for analyzing scientific data. *Science*, 229(4716):828–833, 1985. 4, 9

Victor Dibia and Çagatay Demiralp. Data2vis: Automatic generation of data visualizations using sequence-to-sequence neural networks. *IEEE Transactions on Visualization and Computer Graphics*, 2019. 1, 2, 3

Kushal Kafle, Brian Price, Scott Cohen, and Christopher Kanan. DVQA: Understanding data visualizations via question answering. In *Proceedings of the IEEE/CVF Conference on Computer Vision and Pattern Recognition (CVPR)*, pp. 5648–5656, 2018. doi: 10.1109/CVPR.2018.00592. 3

Samira Ebrahimi Kahou, Vincent Michalski, Adam Atkinson, Ákos Kádár, Adam Trischler, and Yoshua Bengio. FigureQA: An annotated figure dataset for visual reasoning. In *ICLR Workshop on Learning to Generate and Understand*, 2018. URL https://arxiv.org/abs/1710.07300. 3

Shankar Kantharaj, Rixie Tiffany Leong, Xiang Lin, Ahmed Masry, Megh Thakkar, Enamul Hoque, and Shafiq Joty. Chart-to-text: A large-scale benchmark for chart summarization. In Smaranda Muresan, Preslav Nakov, and Aline Villavicencio (eds.), *Proceedings of the 60th Annual Meeting of the Association for Computational Linguistics (Volume 1: Long Papers)*, pp. 4005–4023, Dublin, Ireland, May 2022. Association for Computational Linguistics. doi: 10.18653/v1/2022.acl-long.277. URL https://aclanthology.org/2022.acl-long.277. 2

Jae Kim, Lisa Park, and Rui Wang. Grounding visual language models for chart understanding. In *Proceedings of the IEEE Conference on Computer Vision and Pattern Recognition*, pp. 1234–1243, 2023. 1

Haotian Liu, Pengchuan Jen, Luowei Zhang, et al. Visual instruction tuning of large language models. In *Advances in Neural Information Processing Systems*, 2023. 1, 2

Chen Ma, Hong Ren, and Benjamin Bach. Adavis: Adaptive visualization recommendation through learned design patterns. In *Proceedings of the ACM Conference on Human Factors in Computing Systems*, 2021. 1, 2, 3

Jock Mackinlay. Automating the design of graphical presentations of relational information. *ACM Trans. Graph.*, 5(2):110–141, April 1986a. ISSN 0730-0301. doi: 10.1145/22949.22950. URL https://doi.org/10.1145/22949.22950. 2

Jock Mackinlay, Pat Hanrahan, and Chris Stolte. Show me: Automatic presentation for visual analysis. *IEEE transactions on visualization and computer graphics*, 13(6):1137–1144, 2007. 2

Jock D. Mackinlay. Automating the design of graphical presentations of relational information. *ACM Transactions on Graphics*, 5(2):110–141, 1986b. 4, 9

Paula Maddigan and Teo Susnjak. Chat2vis: Generating data visualizations via natural language using chatgpt, codex and gpt-3 large language models. *Ieee Access*, 11:45181–45193, 2023. 3

Ahmed Masry, Xuan Long Do, Jia Qing Tan, Shafiq Joty, and Enamul Hoque. Chartqa: A benchmark for question answering about charts with visual and logical reasoning. In *Findings of the association for computational linguistics: ACL 2022*, pp. 2263–2279, 2022. 3

Arpit Narechania, Arjun Srinivasan, and John Stasko. Nl4dv: A toolkit for generating analytic specifications for data visualization from natural language queries. *IEEE Transactions on Visualization and Computer Graphics*, 27(2):369–379, 2020. 3

Pew Research Center. Pew research center visualizations. https://www.pewresearch.org/, 2015. Accessed 2025-10-26. 3

Steven F Roth, John Kolojejchick, Joe Mattis, and Jade Goldstein. Interactive graphic design using automatic presentation knowledge. In *Proceedings of the SIGCHI conference on Human factors in computing systems*, pp. 112–117, 1994. 2

Rohan Sharma, Feng Li, and Ning Xu. Plotgen: Generating matplotlib code from natural language descriptions. In *NeurIPS Workshop on Data Visualization*, 2023. 5

Benny Tang, Angie Boggust, and Arvind Satyanarayan. VisText: A benchmark for semantically rich chart captioning. In Anna Rogers, Jordan Boyd-Graber, and Naoaki Okazaki (eds.), *Proceedings of the 61st Annual Meeting of the Association for Computational Linguistics (Volume 1: Long Papers)*, pp. 7268–7298, Toronto, Canada, July 2023. Association for Computational Linguistics. doi: 10.18653/v1/2023.acl-long.401. URL https://aclanthology.org/2023.acl-long.401/. 3

Yuan Tian, Weiwei Cui, Dazhen Deng, Xinjing Yi, Yurun Yang, Haidong Zhang, and Yingcai Wu. Chartgpt: Leveraging llms to generate charts from abstract natural language. *IEEE Transactions on Visualization and Computer Graphics*, 31(3):1731–1745, 2024. 2, 3

Weilin Wang, Yujia Ding, and Yang Chen. Datasetnet: An annotated dataset for data extraction from statistical charts. In *International Conference on Pattern Recognition*, 2020. 2

Zirui Wang, Mengzhou Xia, Luxi He, Howard Chen, Yitao Liu, Richard Zhu, Kaiqu Liang, Xindi Wu, Haotian Liu, Sadhika Malladi, Alexis Chevalier, Sanjeev Arora, and Danqi Chen. Charxiv: Charting gaps in realistic chart understanding in multimodal llms. *arXiv preprint arXiv:2406.18521*, 2024. 2, 3

Yuki Watanabe, Michael Johnson, and John Kelleher. Visml: Learning aesthetic design rules for automated visualization recommendation. In *IEEE Visualization*, 2022. 2

Kanit Wongsuphasawat, Leilani Battle, Arvind Srinivasan, et al. Voyager: Exploratory data analysis via recommendation. In *Proceedings of the ACM Conference on Human Factors in Computing Systems*, 2016. 2

## Ethics Statement

This work trains models on publicly available chart datasets (PewResearch and CharXiV) and uses an LLM as an automated evaluator. No personal data is collected. The LLM judge introduces potential evaluation bias, mitigated via human validation (92% agreement). The system is designed for assistive data visualization and poses no identified misuse risks.

# A Appendix

### A.1 Prompt Templates

**CSV Extraction Prompt**

You are a chart understanding model. Carefully read the following chart image `<image_1>`. Extract the data shown in the chart as CSV format. If the image contains multiple charts, output each as `Chart 1: ...csv...`, `Chart 2: ...csv...`. Include column headers. Output only CSV. No extra text.

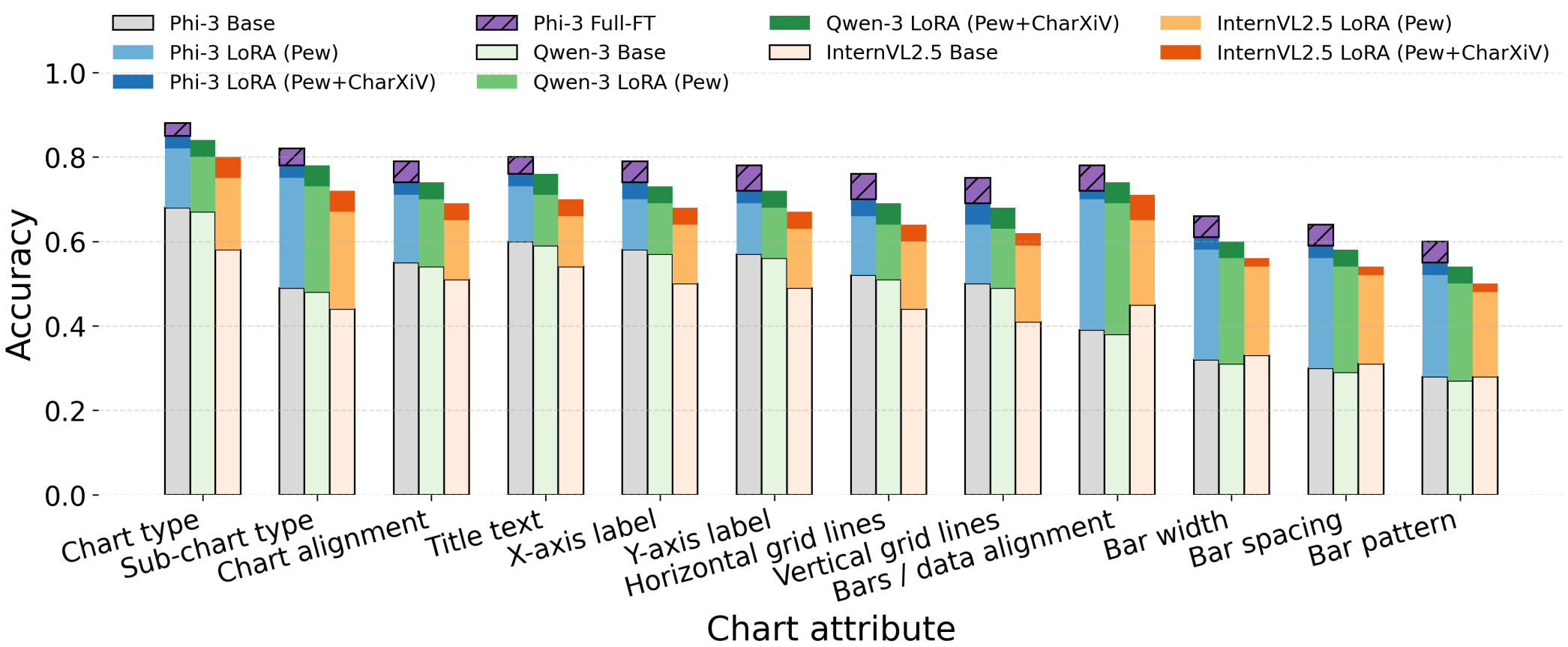

Figure 7: **Attribute-wise accuracy on the PewResearch test set.** Comparison of base models, LoRA fine-tuning, and full fine-tuning across different chart attributes.

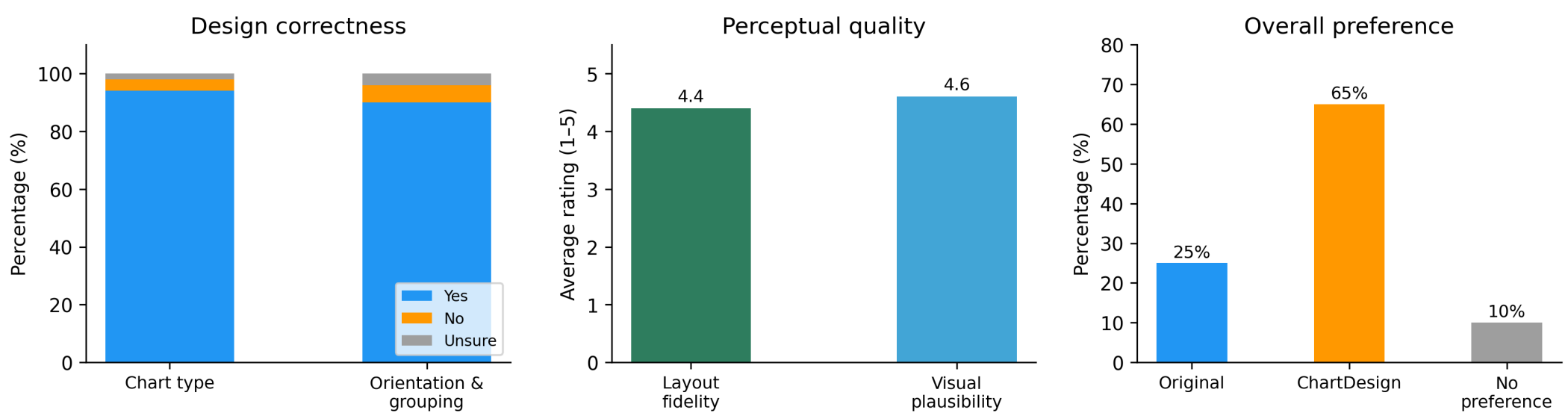

Figure 8: **Summary of human evaluation results.** Left: distribution of responses for chart type correctness and orientation/grouping questions. Centre: average ratings for layout fidelity and visual plausibility. Right: annotators' overall preferences for the original vs. predicted charts.

**Design Choice Prompt**

You are a vision-language model that analyzes chart images and their associated CSV data. Given the chart image `<image_1>` and the CSV data below, generate a JSON that describes the chart's structure and layout. Follow the specified format. Only include keys that are relevant for the given chart type. Do not include keys with values like "not applicable." If a key does not apply, simply leave it out.

**Evaluation Prompt**

You are evaluating the accuracy of predicted chart design attributes. For each pair of ground-truth and predicted values, respond with `MATCH` if they are semantically equivalent (e.g., "bar" and "bar chart") and `NO_MATCH` otherwise. Use natural-language reasoning to decide semantic equivalence.

### A.2 Training details

To adapt compact language models to the chart design task we employ parameter-efficient fine-tuning with LoRA adapters. We freeze all pretrained weights of Phi-3 mini (4.2 B parameters) and Qwen-3 (8 B) and insert rank-16 low-rank projections into each attention and feed-forward layer. Only these adapter weights are updated during training. Each training sample comprises a two-message chat: the first message contains a task instruction (e.g., "Generate a JSON for chart attributes") and the second contains the extracted CSV table; the assistant response contains the annotated design JSON. We set maximum sequence lengths of 14 k tokens for Phi-3 and 8 k for Qwen and train for three epochs on 90% of the corpus using AdamW (learning rate $1 \times 10^{-4}$) with mixed-precision.

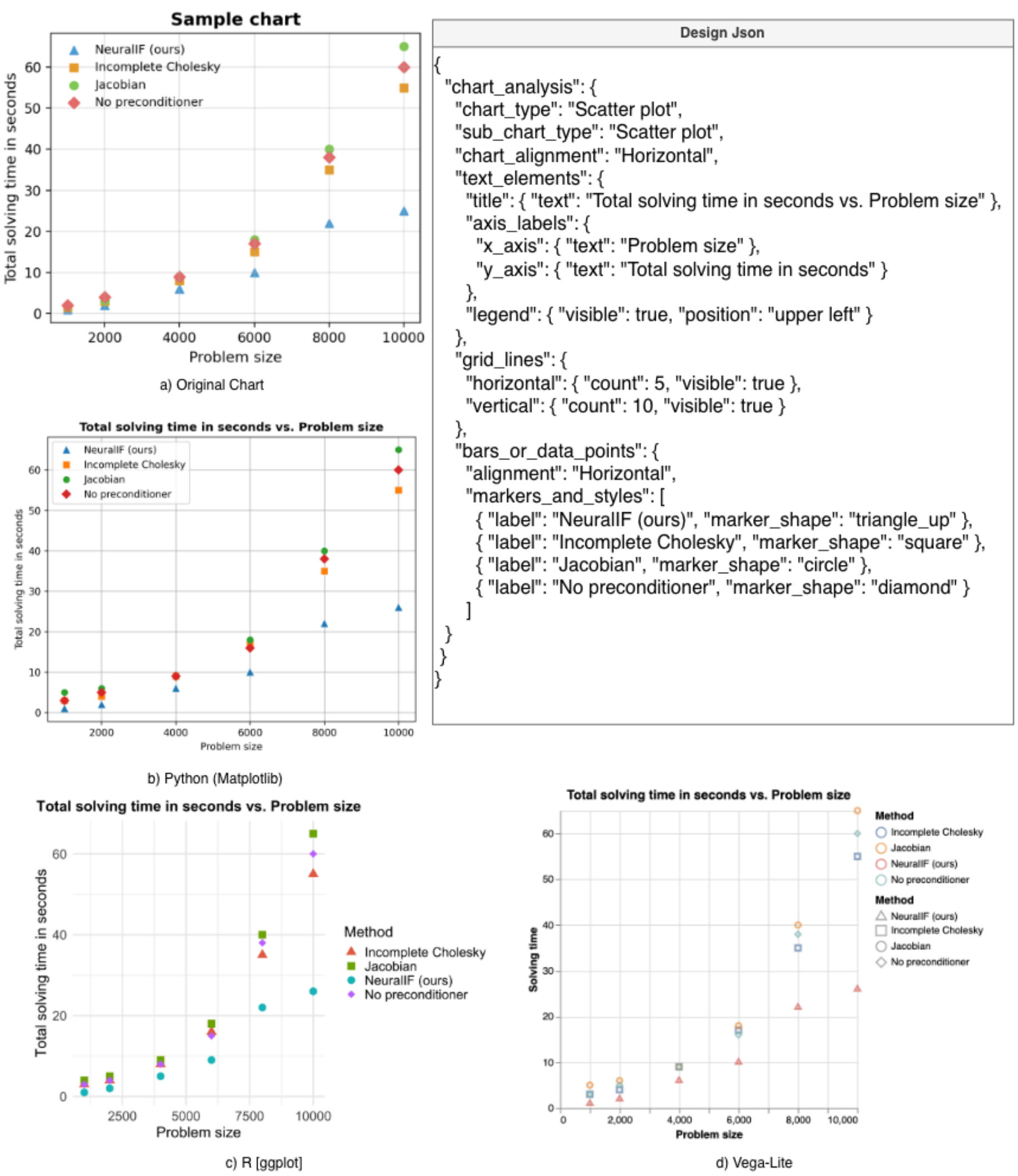

Figure 9: **Language-independent chart specification.** A single structured design JSON enables faithful reproduction of the same visualization across Python (Matplotlib), Vega-Lite, Altair, and R (ggplot2). Each panel uses enlarged fonts for axes and legends to ensure readability after two-column scaling. See Appendix A.5 for a full description of the design schema.

Because many design attributes are imbalanced, we draw mini-batches using an inverse-frequency sampling scheme. Let $N$ denote the total number of training examples and $n_k$ the number of examples containing attribute value $k$. We assign an importance weight $w_k = \log(1 + N/n_k)$ to each value $k$ and weight each example by the product of its applicable $w_k$ (with a small penalty for missing attributes). Batches of four examples are sampled with replacement according to these weights, which increases the frequency of rare attributes during training.

### A.3 Evaluation details

Evaluation is based on per-attribute and overall accuracies on the manually verified test sets. We flatten each predicted and ground-truth JSON into attribute-value pairs and align them by attribute path. A separate LLM judge (prompted as in Appendix A.1) determines whether each predicted value is semantically equivalent to the reference. Each attribute is scored as `MATCH` or `NO_MATCH`, and we report the proportion of matches as the attribute-wise accuracy. Averaging over all attributes yields the overall accuracy reported in the main paper.

### A.4 Dataset statistics

Table 2 in the main paper summarises the composition of our training corpus across the PewResearch and CharXiV sources. Here we provide additional clarification on how these statistics inform model training. Counts correspond to the *primary* chart type per image, while grouped and stacked bars are treated as sub-types of bar charts. The diversity in chart alignment, grid-line density, bar width, spacing, and style patterns directly motivates the inverse-frequency sampling strategy described in Appendix A.2, ensuring that rare design attributes are sufficiently represented during fine-tuning.

### A.5 Design annotation schema

Each chart is annotated with a hierarchical JSON that captures high-level chart families and fine-grained stylistic choices. The top-level field `chart_type` specifies the general family (bar, line, area, scatter, pie, boxplot); `sub_chart_type` encodes variations such as grouped or stacked bars. The `chart_alignment` field indicates whether marks are oriented horizontally or vertically. Free-form strings for titles and axis labels are stored in the `text_elements` object, and the `axes` object records categorical categories or numeric ranges for the x- and y-axes. A `legend` entry controls visibility, position and label text when multiple series are present.

For bar- and box-type charts, the `bars_or_data_points` object specifies the alignment of graphical marks (grouped vs. stacked), their width and spacing, and a pattern property (solid, striped, dotted). The `grid_lines` field counts horizontal and vertical grid lines to encode approximate scale. Additional objects such as `size_and_spacing` and `boxplot_style` capture box widths, intra-group spacing and box-plot-specific parameters (whisker rule, outlier marker visibility, mean markers). When an attribute does not apply (e.g., a pie chart has no axes) the corresponding key is omitted. Figure 9 illustrates how a single design JSON can be rendered across Matplotlib, Vega-Lite, Altair and ggplot2.

### A.6 Ablation details

We evaluate the impact of inverse-frequency sampling and training data diversity. Uniform sampling reduces accuracy on rare attributes (e.g., spacing and pattern) by 5-10 points. Training only on PewResearch lowers performance on CharXiV charts by approximately 7 points, confirming the importance of balanced sampling and cross-domain data.

### A.7 Human study details

During our human evaluation (§4.3) ten annotators were each presented with 40 randomly selected chart pairs the original figure and the same data rendered using our predicted design JSON—and answered the following questions:

1. **Chart type correctness**: Does the predicted chart use the same chart family (bar, line, scatter, area, box or pie) as the original? (responses: Yes/No/Unsure)
2. **Orientation and grouping**: Are the orientation (vertical vs. horizontal) and grouping or stacking of elements consistent with the original? (responses: Yes/No/Unsure)
3. **Layout fidelity**: How closely does the predicted chart match the original layout (titles, axis labels, legend placement and spacing)? (Likert scale: 1 = not at all, 5 = identical)
4. **Visual plausibility**: Does the predicted chart look visually plausible and free of obvious rendering errors such as overlaps or truncation? (Likert scale: 1 = poor, 5 = excellent)
5. **Overall preference**: Which chart better communicates the main information or patterns of the data? (options: Original/Predicted/Both equally)

Annotators answered these questions for a random subset of charts covering bar, line, scatter, area and box plots across both PewResearch and CharXiV domains. Table 5 in Appendix A.10 reports aggregated results, and Figure 8 visualises the response distributions.

### A.8 Reproducibility and compute

We will release the following artifacts upon publication: (i) the full dataset (CSV-design JSON pairs) with train/validation/test splits, (ii) data processing and annotation scripts, (iii) inference and evaluation code including the LLM-judge prompts, and (iv) LoRA adapter weights for Phi-3 and Qwen-3 fine-tuned on the combined corpus. All code will be released under the Apache 2.0 licence; chart images from PewResearch and CharXiV remain subject to their original licences.

Table 4: Compute requirements. All experiments used a single NVIDIA RTX A6000 (48 GB VRAM).

| **Task** | **Compute** |
|---|---|
| LoRA fine-tuning (Phi-3, 2,118 charts) | $\approx$6 GPU-hours |
| Full fine-tuning (Phi-3, 2,118 charts) | $\approx$24 GPU-hours |
| Inference (per chart) | $<$3 s (100 input / 50 output tokens) |
| Few-shot prompting (3-shot, per chart) | $\approx$8-12 s (longer context) |

### A.9 Use of AI Assistance

AI-assisted tools were used for limited language editing and literature search. A vision-language model (Phi-3.5-Vision Instruct) was used to extract CSV data and generate initial design annotation JSONs for model training, as described in Section 3. An LLM was also used as an automated judge to evaluate predicted chart design attributes against ground truth (see Appendix A.3). All technical content, experimental design, analysis, and writing decisions were made by the authors.

### A.10 Human study results

Table 5: Human study results (10 annotators, 40 chart pairs). Type and orientation report the percentage of "Yes" responses; layout fidelity and visual plausibility are rated on a 1-5 Likert scale; preference indicates annotators' overall choice between original and `ChartDesign`-generated charts.

| | **Correctness (% Yes)** | | **Rating (mean±std)** | | **Preference (%)** | | |
|---|---|---|---|---|---|---|---|
| | Type | Orient. | Layout | Visual | Orig. | `ChartDesign` | None |
| All charts | 94 | 90 | 4.4±0.5 | 4.6±0.4 | 25 | 65 | 10 |

### A.11 Rare-type generalisation

To assess generalisation to underrepresented chart families, we evaluate on a rare-type subset consisting of scatter, pie, box and histogram charts drawn from each test set. All experiments use the Qwen-3 (8B) backbone; LoRA rows are fine-tuned on the stated corpus, and zero-shot rows use the same backbone without fine-tuning.

Table 6: Rare-type accuracy (%) by training configuration (Qwen-3 backbone).

| **Training Config.** | **Pew Rare ↑** | **Mixed Rare ↑** |
|---|---|---|
| Zero-shot (off-the-shelf) | ≈37 | ≈37 |
| `ChartDesign` LoRA on Pew | 63 | 61 |
| `ChartDesign` LoRA on CharXiV | 65 | 64 |
| `ChartDesign` LoRA on Pew+CharXiV | **78-80** | **78-80** |

Table 6 shows that even with inverse-frequency sampling, zero-shot models achieve only ≈37% on rare types. Fine-tuning on either corpus alone substantially improves performance (63-65%), and combining both domains yields the highest rare-type accuracy of 78-80%, more than doubling the zero-shot baseline. This confirms that multi-domain training and balanced sampling jointly enable generalisation to minority chart families.

### A.12 Error analysis

We systematically analyse the 16% of predictions that fail on our best model (Qwen-3 LoRA on Pew+CharXiV). Table 7 categorises the failure modes by attribute group.

Table 7: Error taxonomy for the 16% of failures on the Pew+CharXiV test set.

| **Error Category** | **Share (%)** | **Description** |
|---|---|---|
| Incorrect sub chart type | 23 | Stacked vs. grouped bar confusion |
| Improper bar spacing | 18 | Bars too close/far from reference |
| Misaligned gridline counts | 15 | H/V gridline count mismatch |
| Legend placement / label order | 12 | Entries or axis labels mis-ordered |
| Width / height ratio | 10 | Aspect ratio differs from reference |
| Other attribute mismatches | 22 | Minor spacing or styling differences |

The largest error class (23%) involves sub-chart-type confusion, predominantly between stacked and grouped bars-variants that differ in a single JSON field. Bar spacing and gridline counts reflect continuous value prediction difficulty. Future work could address these via constrained decoding or post-hoc calibration.

### A.13 Robustness to noisy inputs

Real-world CSV data is often imperfect. We evaluate robustness by introducing three perturbation types: (i) random missing cells (up to 10%), (ii) outlier noise (20% of values replaced by random numbers), and (iii) irregular CSV format (missing column headers, extra blank lines). Table 8 reports the accuracy drop relative to clean inputs.

Table 8: Accuracy drop (percentage points) under input perturbations.

| **Perturbation** | **`ChartDesign` LoRA ↓** | **Zero-shot ↓** |
|---|---|---|
| Missing values (≤10% cells) | ≈7 | ≈15 |
| Outlier noise (20% noise) | ≈7 | ≈15 |
| Irregular CSV format | ≈9 | ≈18 |

`ChartDesign` degrades by only 7-9 points under all three perturbations, compared to 15-18 points for zero-shot baselines. Fine-tuning on diverse data builds robustness to realistic input variation, while zero-shot models are more brittle to formatting anomalies and missing numeric context.

### A.14 Masking ablation: data vs. metadata

To understand what signal drives design prediction, we mask numerical cell values in each CSV and retain only column names and axis titles. Table 9 shows that high-level attributes (chart type, orientation) can still be predicted at ≈56% from metadata alone, while fine-grained attributes (bar spacing, pattern) drop sharply to ≈25%.

Table 9: Accuracy (%) with and without numeric data (Qwen-3 LoRA on Pew+CharXiV).

| **Prediction Target** | **Data Masked** | **Full Data** |
|---|---|---|
| Chart type & orientation | ≈56 | 92 |
| Bar spacing / pattern | ≈25 | 82 |

Coarse chart-family decisions are strongly indicated by column semantics (e.g., a date column favours a line chart), while fine-grained layout parameters require access to the actual data distribution.

### A.15 Limitations

**Scope of chart types.** Our design schema focuses on common static 2D chart families (bar, line, scatter, area, pie, box). More complex visualizations such as maps, network diagrams, interactive dashboards, or multi-view coordinated layouts are not currently supported.

**Dependence on data quality.** The approach assumes clean and well-structured tabular input. Errors in data extraction or ambiguous column semantics can propagate to suboptimal design predictions.

**Evaluation limitations.** Our human study involves ten annotators and forty chart pairs; while sufficient to demonstrate statistically meaningful preference trends, larger-scale crowdsourced evaluations would further strengthen conclusions. The current evaluation treats ground-truth human-authored charts as the only acceptable reference, which can be overly strict given that multiple valid designs often exist for a single dataset. Future work should explore multi-reference or preference-based evaluation protocols.

**Scale and model coverage.** Our evaluation focuses on open-source models up to 8B parameters. The impact of fine-tuning on larger models (14B, 70B) and proprietary systems (e.g. GPT-4, Gemini) remains an open question. We expect that larger models benefit less from fine-tuning on a 2k-chart corpus due to stronger zero-shot priors, but this conjecture should be tested empirically.